\pdfoutput=1 % arXiv admin added this line
\documentclass[aps,prb,showpacs,floats,twocolumn,amsfonts,amssymb,unsortedaddress,floatfix]{revtex4}
\input epsf
\usepackage{bm}
\usepackage{amsmath}
\usepackage{graphicx}

\begin{document}

\title{Polaron Induced Deformations  in Carbon Nanotubes}
\author{Cristiano Nisoli}
\affiliation{Theoretical Division and Center for Nonlinear Science, Los Alamos National Laboratory, Los Alamos NM 87545}

\date{\today}
\begin{abstract}
We compute for the first time full elastic deformations, as well as length, of self-trapped electronic states in carbon nanotubes of general radius and chirality, within the unifying framework of a recently introduced two field model for electromechanics of carbon nano-structures. We find that deformations are highly non monotonic in the chiral angle, whereas the  length of the polaron is  not. Applications include nano-mechanical devices as electrically or optically driven nano-actuators. 
 \end{abstract}


\maketitle

Coupling between electronic and elastic structure  of carbon nanotubes~\cite{ji, imperial, Dresselhaus} leads to  theoretically interesting and technologically relevant phenomena: among other things,  gap opening can be induced via strain~\cite{Rochefort,Yang}  (allowing mechanical control of the conductance) and injecting electron/holes induces mechanical deformations~\cite{Verissimo, Gartstein} (important for realization of electrical nano-actuators)  or shifts in Raman phonon modes~\cite{Margine,Chen1,Chen2}.  
Novel excitations leading to spontaneous distortions driven by electron-phonon couplings are of particular interest to the electromechanics of nano devices ~\cite{Lammert, Chamon}. In particular light induced mechanical deformations, or optomechanical effects can be observed at high temperatures in nanotube, as excitons are more strongly bound~\cite{Dresselhaus2}. Recently,  Verissimo-Alves {\it et al}~\cite{Verissimo0} have predicted the existence of self-trapped electron and hole states in semiconducting carbon nanotubes by feeding parameters from {\it ab initio} density functional calculations  into  a very simplified continuum model. They found lengths of the order of 40-60 nm and energies of order of $10^{-2}-10^{-1}$ meV. Their long polaron  approximation is confirmed by a  atomistic numerical analysis e.g. of Bratek {\it et al}~\cite{Bratek}.

Chiral indices are known to play dramatic role in the physics of carbon nanotubes. Not only they dictate the metallicity, they are known to control in a very delicate way electromechanical and vibrational effects~\cite{Margine, Chen1, Chen2}.  It is of outmost importance both theoretically and technologically to be able to predict  for which chirality a polaron induces e.g. contraction rather than elongation, radial expansion rather than shrinking,  or torsion et cetera. While Ref.~\cite{Verissimo0} first suggested the presence of polarons, their simplified model cannot address the complex  deformations in carbon nanotubes and completely ignores chirality dependence. On the other hand, more complete treatments based on atomistic models  are too complex to be solved analytically or too computationally costly to return detailed answers for a general case. 

In this letter we employ a recently introduced~\cite{Nisoli} bicontinuum model to solve this predicament and  calculate the full range of deformations induced by  self-trapped electrons/holes for the general carbon nanotube. We find that deformations are, as one would expect, highly non monotonic in the chiral indices,  even with abrupt change of signs in nanotube of the same chiral index and about the same radius.

\begin{figure}[t!]
\center
\vspace{1 mm}
\includegraphics[width=2.2 in]{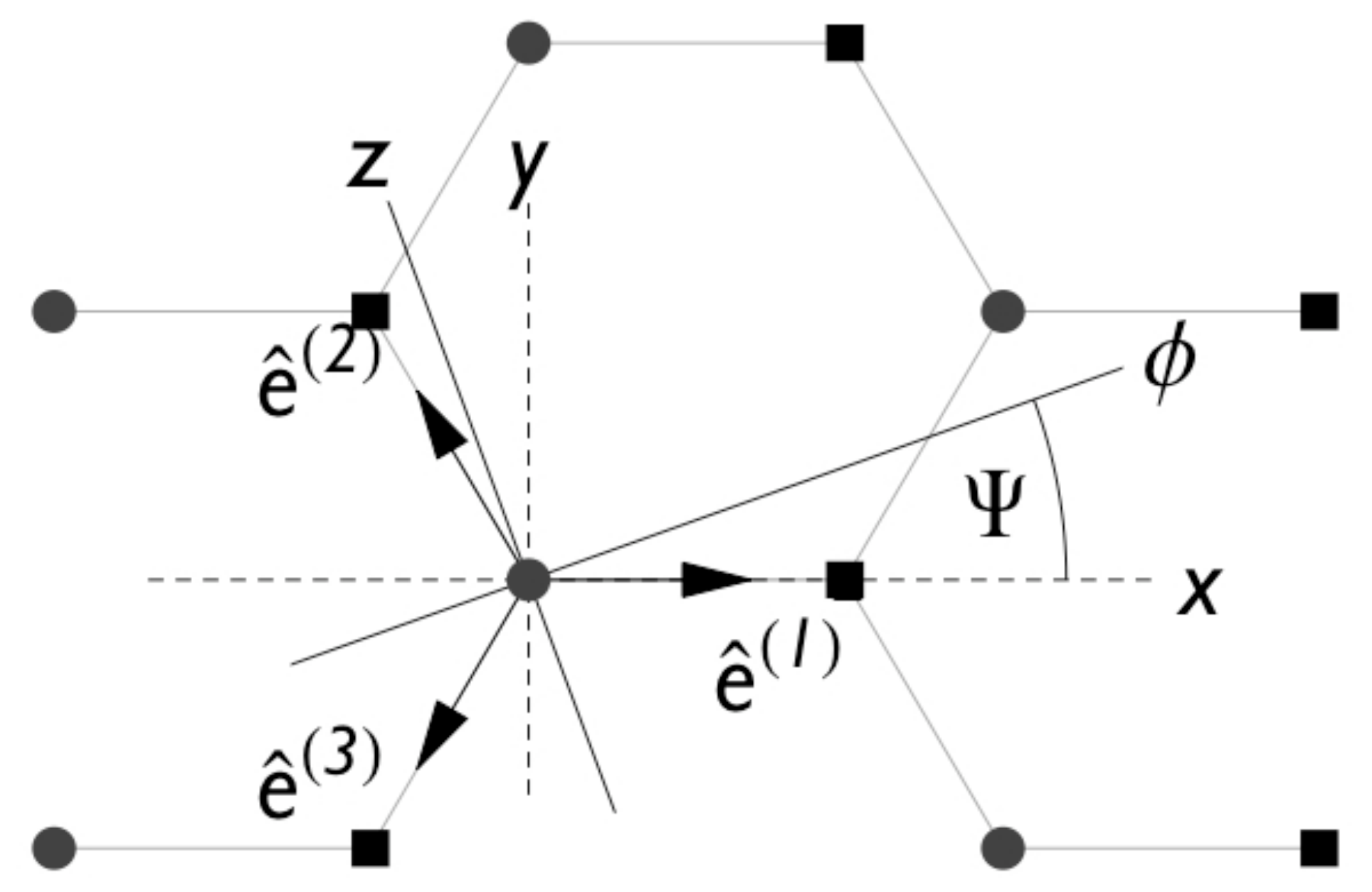}
\caption{The two sublattices (circles and squares) of graphene and the three unit vectors $\hat{e}^{(l)}$ used in the text. ${\phi}$, ${z}$ are cylindrical coordinates of a tube, while $\Psi=\pi/6-\theta_c$ with $\theta_c$ the chiral angle. }
\label{lattice}
\end{figure}

 In a recent work, Nisoli {\it et al}~\cite{Nisoli} have introduced a two field unifying framework for elasticity, lattice dynamics and electromechanical coupling in carbon nanostructures that  accounts for the full atomistic detail of the graphenic lattice, and  explains a wealth of experimental and numerical results  without computationally intensive atomistic treatments.  They defined an elastic field for each of the mutually interlaced triangular sublattices that make the honeycomb lattice, $u^i(x)$, $v^i(x)$, $i=1,2$ (see Fig.~\ref{lattice}) and the corresponding strain tensors~\cite{Landau}  $u^{ij}=\partial^{(i}u^{j)} $,
$v^{ij}= \partial^{(j}v^{i) }$, and wrote the elastic energy via considerations of symmetry.  Unlike Ref.~\cite{Nisoli}, we work here with  the average displacement $2 p^i=(u^i+v^i)$, inner displacement $ 2 q^i=(u^i-v^i)$, and corresponding strain tensors $p^{ij}$, $q^{ij}$.  In these new variables, the elastic energy of  Ref.~\cite{Nisoli} becomes
%for the kinetic energy  $T=\frac{1}{2}\int \sigma \left(\dot{p}^2+\dot{q}^2\right) \mathrm{d} x^2$ while 
 $
W_{{\rm el}}=\int \sigma\!\ V_{{\rm el}}[p,q]\!\ \mathrm{d}x^2$, with
\begin{eqnarray} 
V_{{\rm el}}[p,q]&=&\hat\mu \!\ p^{ij}p_{ij}+\frac{\hat \lambda}{2} \!\ p^i_i p^j_j  \nonumber \\ 
&+&\frac{1}{2}\omega_{\Gamma}^2 \!\ q^2 -2\beta \!\ e_{ijk}q^i p^{jk},
\label{Vpq}
\end{eqnarray}
where we have employed the long polaron approximation to neglect the dispersion of the optical branches
\footnote{The expression for $V_{{\rm el}}$ is simplified in the long polaron approximation. Using the notation of Ref.~\cite{Nisoli}, we can neglect terms in $\partial^i q^j$ (the dispersion in the optical branches) as long as $(\mu-\mu')/a^2$,  $(\lambda-\lambda')/a^2$ $\ll \alpha$, where $a$ is the length of the polaron. Since  $(\mu-\mu')$,  $(\lambda-\lambda')$ are the order of the square of the speed(s) of sound in graphene\cite{Nisoli} or about $10^8$ m$^2$s$^{-2}$, whereas $\alpha$ can be expressed in terms of the graphite-like optical frequency\cite{Nisoli} $\alpha=\omega_{\Gamma}^2/4\sim10^{28}$ s$^{-2}$ and we have for the length of the polaron\cite{Verissimo0}   $a\sim 10$ nm, those ratios are thus of the order of 
%$ c^2/a^2\omega_{\Gamma}^2\sim 
$10^{-4}$. 
}. 
 The tensor $e_{ijk}$ is invariant under the $C_{3v}$ group and can
be represented by the three unit vectors $\{\hat{e}^{(l)}\}_{l=1,3}$ of
Fig.~\ref{lattice}, 
\begin{equation} 
e_{ijk}=\frac{4}{3} \sum_{l=1}^3 \hat{e}^{(l)}_i \hat{e}^{(l)}_j \hat{e}^{(l)}_k. 
\label{B}
\end{equation} 
There are thus 4 parameters:
 $\omega_{\Gamma}$,  the graphite-like optical frequency; $\beta$ which determines the strength of the rotational symmetry breaking,  contains all information about the point group symmetry of
graphene, and defines the important length $\ell \equiv 4 \beta/\omega_{\Gamma}^2 =0.3$~\cite{Nisoli}; and the generalized Lam\`e symbols $\hat \mu$, $\hat \lambda$,  which can be expressed in terms of the longitudinal and transverse speed of sound in graphene, $v_{\rm L}^2=2\hat\mu+\hat\lambda-4\beta^2/\omega_{\Gamma}^2$, $v_{\rm T}^2=\hat\mu-4\beta^2/\omega_{\Gamma}^2$, and are related to the actual Lam\`e symbols of graphene $\mu_r$, $\lambda_r$ via  $\mu_{\mathrm{r}}=\hat\mu-4\beta^2/\omega_{\Gamma}^2$ and $\lambda_{\mathrm{r}}=\hat\lambda+4\beta^2/\omega_{\Gamma}^2$. We then write $p$ in term of the isotropic ($\gamma_o$), anisotropic ($\gamma$), and shear/torsional ($\eta$) strain in the nanotube coordinates 
\begin{equation}
\left\{ \begin{array}{l} 
p^{\phi\phi}=\gamma_o+\gamma\\
p^{zz}=\gamma_o-\gamma\\
p^{\phi z}=p^{z \phi}=\eta
\end{array} \right . .
\label{strain}
\end{equation}
In these new variables the elastic energy  (\ref{Vpq}) now reads
\begin{eqnarray} 
V_{{\rm el}}&=&2\hat \lambda \!\ \gamma_o^2 +2\hat\mu \left(\gamma_o^2+\gamma^2+\eta^2\right)+\frac{1}{2}\omega_{\Gamma}^2 \!\ q^2  \nonumber \\ 
&-& 4\beta \!\ q^{\phi}\left(\gamma s_3-\eta c_3\right)+4\beta \!\  q^{z}\left(\gamma c_3+\eta s_3\right).
\label{Vgq}
\end{eqnarray}
(We have shortened $c_3\equiv \cos (3\theta_c)$, $s_3\equiv \sin (3\theta_c)$, $\theta_c $ is the chiral angle of the nanotube, and  used  $ e_{\phi,\phi,\phi}= - e_{\phi,z,z}=-s_3 $, and  $e_{z,z,z}= - e_{\phi,\phi,z}=-c_3$.)

As explained in Ref.~\cite{Nisoli}, one can simply ``wrap around'' the elastic energy of graphene to deal with carbon nanotubes. In the cylindrical geometry, with coordinates
$\{r,\phi, z\}$ of Fig.~\ref{lattice}, a minimal coupling between  the tangential displacements $p^i$ and the radial  $p^r$ appears in $V_{{\rm el}}$ of (\ref{Vpq}) via
$
p^{\phi\phi}=\left(\partial_{\phi}p^{\phi}+p^r\right)/r
$~\cite{Landau}.  As explained, we assume no azimuthal dependence, and thus $2 \gamma_o=\left(\partial_z p_z+p_{\mathrm{r}}/r\right)$, $2 \gamma=\left(-\partial_z p_z+p_{\mathrm{r}}/r\right)$, $2 \eta=\partial_zp_{\phi}
$.
We are neglecting  the  breaking of  the hexagonal symmetry brought upon by the chiral vector that defines the wrapping of the carbon nanotube. This symmetry breaking  {\it allows for new terms to be introduced in $V_{{\rm el}}$ as  curvature corrections}. 
%These terms can be obtained by constructing  with $u$, $v$ tensors respective of the hexagonal symmetry (see the end of the section) and contracting those with the  vector $z^i$  of Fig.~\ref{lattice}. We shall here neglect  all these corrections, although the reader should be cautioned: as we shall se later, polarons are more evident  in small radius nanotubes.
The parameters of our elastic energy for graphene are also corrected by curvature. 

As for the  low energy electronic excitations in a semiconducting nanotube, they can be described in terms of an envelope wave-function $\psi$~\cite{Madelung} of energy density
\begin{equation}
E_e[\psi,p,q]=-\psi^{\dag}\frac{\hbar^2\partial_z^2}{2 m} \psi + E_{{\rm ep}}[\psi,p,q].
\label{Ee}
\end{equation}
$E_{{\rm ep}}$, the coupling between phonons and an injected electron or hole,
at lowest order  both in the  {\it in-plane} elastic fields  and in the electron probability density $|\psi|^2$ 
can be deduced via considerations of symmetry 
\begin{eqnarray}
E_{{\rm ep}} &=&  |\psi|^2 \left[\nu\left(q^z/e+\gamma
    c_3 +\eta s_3 \right) +\nu' \gamma_o\right]
\label{Ep}
\end{eqnarray} 
($e$ is the bond length).  Physically, the term proportional to $\nu$ emerges in dielectric tubes from a deformation-induced change in the bandgap, whereas the term proportional to $\nu'$ comes from the shift in energy at the K point of the Brillouin due to second-nearest-neighbor atoms~\cite{Verissimo}. 
%The out of plane inner displacement $q^r$ does not appear in (\ref{Ep}) because it neither breaks the hexagonal symmetry of graphene and thus cannot change the band gap in dielectric carbon nanotubes, nor can shift the overall energy as the K point, since it connects nearest neighboring atoms.  
In the context of a simple tight-binding treatment~\cite{imperial},  hopping integrals are modulated by in-plane elastic deformations.~\cite{Nisoli}: $\mathrm{d}t^{(l)}= - \tau \!\ \hat{e}^{(l)}_i \hat{e}^{(l)}_j p^{ij} +\tau \!\ \hat{e}^{(l)}_i q^i/e$ are the three nearest-neighbor hopping integrals along the three bonds of unit vectors $\hat{e}^{(l)}$, whereas 
$\mathrm{d}t'^{(l)}= - \tau' \!\ \hat{a}^{(l)}_i \hat{a}^{(l)}_j p^{ij}$ are the hopping integrals along the three directions $\hat{a}^{(l)}$ of the next-nearest-neighbors. ($\tau$, $\tau'$ are often called scaling parameters~\cite{Harrison}.) Following along the line of Ref.~\cite{Nisoli,Yang}, one can calculate the variation of the band gap and of the energy at the K point under strain via hopping integral modulation, and---after long yet not particularly insightful calculations---deduce (\ref{Ep})  and   in particular
\begin{equation}
\left\{ \begin{array}{l} 
\nu'=\mp 3 \tau' \\
\nu=\frac{3}{2} s_p \tau.
\end{array} \right . 
\label{nu'}
\end{equation}
The minus sign in the first equation is for electrons, the plus for holes.  Note that there is no change in sign for $\nu$ in going from holes to electrons. The other sign function is $s_1= -1$ ($s_2=1$) where $p$ is defined as a function of $n$, $m$ as  $p\equiv [(n-m)\mod3]$.   This difference in sign behavior between $\nu$ and $\nu'$ was previously recognized  as causing indices dependence and non monotonicity in the doping induced shift of Raman frequencies and anomalous bond contraction/expansion in zig-zag nanotubes~\cite{Nisoli,Margine, Chen1,Chen2}. We will see that it controls chirality dependent  expansions vs. contractions for hole or electron self-trapping.

We assume that our fields only vary along the axial coordinate $z$, an ansatz corroborated by previous numerical calculations, as explained above~\cite{Bratek}. This treatment---we will see---is self consistent as it predicts very long polarons. The  hamiltonian density for the entire system is then
\begin{eqnarray}
{\cal H}=-\psi^{\dag}\frac{\hbar^2\partial_z^2}{2 m} \psi+  \sigma c V_{{\rm el}}+ E_{{\rm ep}}
\end{eqnarray}
where $V_{{\rm el}}$ is given by (\ref{Vgq}),  $E_{{\rm ep}}$ is given by~(\ref{Ep}), and $c=2\pi r$ is the nanotube circumference.  Minimization under normalization of $\psi$ with Lagrangia multiplier $\epsilon_p$ returns the system of equations of equilibrium for our system,
\begin{equation}
\left\{ \begin{array}{l} 
4 (\hat \mu+\hat \lambda ) \!\  \gamma_o =-\nu ' |\psi | ^2/\sigma c  \\ 
4\hat \mu \!\  \gamma -4 \beta (q_{\phi} s_3- q_z c_3)=-\nu c_3 |\psi |^2/\sigma c  \\ 
4\hat \mu \!\  \eta + 4 \beta (q_{\phi} c_3+q_z s_3)=-\nu s_3 |\psi |^2/\sigma c  \\ 
4 \beta (\gamma c_3+ \eta s_3)+\omega_{\Gamma}^2 q_{z} =-\nu |\psi |^2/ e\sigma c  \\
4 \beta (\gamma s_3- \eta c_3)-\omega_{\Gamma}^2 q_{\phi} =0 \\ 
 \epsilon_p \psi=\frac{-\hbar^2}{2 m}\partial_z^2 \psi +\left[\nu\left(q^z/e+\gamma
    c_3 +\eta s_3 \right) +\nu' \gamma_o\right]\psi  \\
\end{array} \right. 
    \label{static}
\end{equation}
which is linear in the elastic fields and can be solved easily for them in terms of $|\psi|^2$. 

We find  for the local deformation induced by an injected electron or hole the equations
\begin{equation}
\left\{ \begin{array}{l} 
\gamma_o=\pm\xi_o \!\ |\psi |^2 \\ 
\gamma=-s_{p} \!\ \xi \cos(3\theta_c) |\psi |^2 \\ 
\eta=-s_p \!\ \xi  \sin(3\theta_c) |\psi |^2 \\ 
\end{array} \right. ,
\label{gamma}
\end{equation}
which show that strain is higher in regions where the electron is localized.  We find $\xi_o=\frac{3\tau'}{4 \!\ \sigma c \!\  (v_{\rm L}^2-v_{\rm T}^2 )}$,  $\xi=\frac{3 \tau  \left(1+{l}/{e}\right)}{8\!\  \sigma c\!\ v_{\rm T}^2} $ inversely proportional to the circumference.  The sign in the first equation is plus for electrons and minus for holes. The other two equations change sign depending on the chirality of the nanotube via $s_p$.  
%The lengths $\xi_o$, $\xi$ read 
%\begin{equation}
%\left\{ \begin{array}{l} 
%\xi_o=\frac{3\tau'}{4 \!\ \sigma c \!\  (v_{\rm L}^2-v_{\rm T}^2 )}\\
%\xi=\frac{3 \tau  \left(1+{l}/{e}\right)}{8\!\  \sigma c\!\ v_{\rm T}^2}
%\end{array} \right. 
%\end{equation}
%and are inversely proportional to the circumference. 
%and $l\equiv 4\beta/\omega_{\Gamma}^2$ is a characteristic length that expresses the coupling between the inner displacement and the strain fields. Nisoli {\it et al} found it to be  $l\simeq 0.2-0.3$~\AA~~\cite{Nisoli}. 

%Before we derive equations for the other elastic fields and $\psi$,  
From (\ref{strain}) we have for the total elongation and torsion of the tube $\Delta L=\int (\gamma_o -\gamma)\!\ {\rm d}z$ and $r \Delta \phi=\int \eta\!\ {\rm d}z$, and thus from~(\ref{gamma}) and because of normalization of $\psi$, we obtain
\begin{equation}
\left\{ \begin{array}{l} 
\Delta L=\pm \xi_o+s_p \!\ \xi \cos(3\theta_c)\\ 
r \Delta \phi=-s_p \!\ \xi \sin(3 \theta_c)
\label{roxana}
\end{array} \right. 
\end{equation}
where, repetita juvant,  the sign in the first equation is plus for electrons and minus for holes. Note that the overall elongation and torsion of the tube due to electron/hole injection is independent of the  actual shape of $\psi$. Unfortunately, that also implies that self-trapped electronic excitations cannot be recognized by global observations such has overall elongation, but  solely  via local mechanical features.
%Further sign dependance is contained in $s_p$ which, as explained above, depends on the chirality indices $n$, $m$. 

Equations (\ref{roxana}) generalize to arbitrary chirality what already found by Nisoli {\it et al} for doping of  zig-zag nanotubes~\cite{Nisoli}. The first result from  (\ref{roxana}) is that electron/hole injection  induces no torsion for zig-zag nanotubes  (which correspond to $m=0$, thus $\theta_c=0$), at least in our approximation that neglects curvature corrections. On the other hand, torsion would be maximum for tubes approaching the armchair configuration (the armchair themselves are metallic and thus excluded from this study): that would be $n=m+1$.  
Again from  (\ref{roxana}) we obtain that elongation is always positive for electrons in nanotubes with $p=2$, whereas is always negative for holes in nanotubes with $p=1$. The maximum in both cases is achieved by zig-zag nanotubes ($\theta_c=0$).  These results are in agreement with the density functional study of Ref.~\cite{Verissimo0}, who found elongation for electrons in (11,0) zig-zag, and contraction for holes in (7,0) zig-zag.
%for maximum elongation is achieved via electron injection in zig-zag tubes for which \mbox{$n \mod 3=2$}. The largest contraction can be achieved by hole injection in zig-zag nanotubes  for which \mbox{$n \mod 3=1$}.

Analysis of the general case requires knowledge of the quantities $\xi_o$,  $\xi$.  If $\xi>\xi_o$, the first equation of~(\ref{roxana}) tells as that chirality can change the sign of $\Delta L$   $s_p$ and thus predicts shortening for self trapped electrons, yet elongation for holes, in certain tubes.  This seems to be the case, from a rough estimate: we use $e=1.42$~\AA  ~and the  Harrison scaling for the hopping parameters~\cite{Harrison}, which are $\tau'=2t'$, $\tau=2 t$, with $t\simeq2.8$ eV, $t'\simeq0.68$ eV~\cite{Reich}. For the speeds of sound in graphene we use  $v_{\rm T}=1.4~10^4$ m s$^{-1}$, $v_{\rm L}=2.16~10^4$ m s$^{-1}$ as in Mahan~\cite{Mahan1}; finally from the density of graphite  2.26 gm cm$^{-3}$ and the interlayer distance at 3.4 \AA~the surface density of graphene can be estimated  $\sigma\simeq7.6~ 10^{-7} {\rm Kg \!\ m^{-2}}$. We obtain $\xi_o\simeq 2.5 \!\ 10^{-2}$ \AA$^2/2r$, $\xi\simeq 8.7 \!\ 10^{-2}$ \AA$^2/2r$. 
% Recent DFT results by Margine {\it et al}~\cite{Margine}  show  that electron injections in zigzag nanotubes always induces lenghtening. Nevertheless, one must bear in mind that numerical simulation of doping via DFT calculations, can only deal with much higher concentration of dopants (limited by the size of the supercell): at those densities,  term quadratic in the density of injected electrons, absent here. order effects come into play. One can easily show that  contributions quadratic in the density of doped electrons are always positive no matter the chirality, as they are related not to the band gap variation, but to the slope of the energy dispersion. The fact that DFT calculations found the lenghtening to be  larger for $p=2$ ($n=14,11$ zig-zag nanotubes)  than for  for $p=1$ ($n=16,13$) is at least a patial confirmation of  (\ref{roxana}).
%For a (7,0) zig-zg nanotube one finds an elongation of 0.025 \AA~ in good agreement with previous results based on DFT calculations~\cite{Verissimo0}. 
For a larger nanotube of 1 nm in diameter, $10^3$ electrons can give a 1nm elongation. Since the elongation and torsion  are inversely proportional to the radius of the nanotube: these effects become stronger for small radii, for which our treatment is only a first order approximation.

%stretching of $b_{ax}$ for  or $n=14,11$ tubes respectively, as predicted by ~\ref{roxana}. In DFT, the overall tube lengthens in the second case , again in accord with the bicontinuum; the lengthening found for $r=2$, is less than for $r=1$, perhaps a consequence of the change in sign in Eqs.~\ref{roxana}. 

From  (\ref{static}) we find that the inner displacement 
\begin{equation}
\left\{ \begin{array}{l} 
q_{\phi}=0 \\
q_z=-s_p \!\ \xi  \left(\l+l' \right)|\psi|^2
\end{array} \right. 
\label{q}
\end{equation}
with  $l'={4 v_{\rm T}^2}/{\omega_{\Gamma}^2(e+l)}\simeq 0.35$~\AA~   (computed from  \mbox{$v_{\rm T}=1.4~10^4$~ms$^{-1}$} and $\omega_{\Gamma}= 3~10^{14}$~s$^{-1}$).  Thus, the inner displacement is always parallel to the nanotube axis, while its absolute magnitude is independent of the chiral angle. On the other hand, its orientation depends on the chiral indices, thus suggesting that polarons can induce highly non monotonic anomalous bond contractions/expansion and thus  hardening/softening of optical modes~\cite{Nisoli,Margine, Chen1,Chen2}.  
%\section{Static Polarons}

Finally,  by substituting the expression for the elastic fields~(\ref{gamma}, \ref{q}) into the last equation of~(\ref{static}), one finds a nonlinear Schr\"oedinger equation for the envelope $\psi$
\begin{equation}
 \epsilon_p  \psi=-\frac{\hbar^2}{2 m}\left(\partial_z^2 \psi  +4 \chi |\psi|^2\psi\right)
\label{NLSE}
\end{equation}
where the reciprocal length $\chi$ is found to be
\begin{equation}
\chi e=\frac{\pi}{4 \sigma c^2 t}\left[ \frac{\tau^2 (e+l)(e+l+l')}{ v_{\rm T}^2 e^2 } + \frac{\tau'^2}{4 (\hat \mu+\hat \lambda ) }\right] 
\label{len}
\end{equation}
where we have used the expression $m=2\pi \hbar^2/(9e c t)$ for the effective mass of the electron  obtained from the formula for the band gap~\cite{imperial}. The well known self-bound solution of (\ref{NLSE})

\begin {equation}
\psi(z,t)=\sqrt{\frac{\chi}{2}} \cosh^{-1}(\chi z)e^{-i\epsilon_p t/ \hbar}
\label{nlse}
\end{equation}
corresponds to a polaron of energy
$\epsilon_p=-\frac{\hbar^2 \chi^{2}}{2 m}$
and  length  $a=\chi^{-1}$.
Remarkably unlike the amplitude of the elastic deformations, {\it the length and thus the energy of the polaron does not depend on the chiral angle}. Also  polaron length {\it is the same for electrons and holes}. This latter statement is in contrast with results reported by Verissimo-Alves~\cite{Verissimo0}: while for the (11,0) nanotube they find about the same polaron length for electron (39 nm) and hole (40 nm),  for the (7,0) case the electron polaron has a length of 59 nm, while the hole is 21 nm long.  We suspect that the discrepancy originates  a sign error in the quantity in Ref~cite{Verissimo0} defines as $\lambda_z$, as analysis of their Fig.~2 confirms. In our formalism, and using the Harrison scaling and parameters above, we have for a zigzag nanotube $\lambda_z=-3s_p t/2\mp3t'$ and estimate $\lambda_z=-6.3$  eV for electrons, -2.2 eV for holes, for a (10,0) nanotube--in the ballpark with the Ref.~\cite{Verissimo0} values of -8.3, -2.4 eV.  Instead we find a sign mismatch for the (7,0) where our estimates are  +2.2 eV for electrons, +6.3 eV for holes,  whereas they have -1.6 eV, +7.8 eV.  The polaron length $a=\chi^{-1}$ can be estimated from~(\ref{len}) to be
%  using the numbers introduced before to be $a\simeq 3 \!\ 10^{2} \left(\frac{t}{\tau} \frac{c}{1{\rm  nm}}\right)^2\simeq 7 \!\ 10^{2} \left(\frac{D}{1{\rm  nm}}\right)^2 $, (as $(t/ \tau)^2\sim 10^{-1}$)  Hence  nanotubes of a nanometer diameter $D$ should show polarons 
of several hundred angstroms for nanotubes of a few nanometers in diameter, in agreement with estimate based on DFT results~\cite{Verissimo0}. 

Finally, for completeness, we calculate the radial variation of the nanotube $\Delta r/r=p^{\phi \phi}$. From (\ref{gamma}) and (\ref{nlse}) we see that  the maximum radial variation corresponds to the maximum of $\psi$ (at $z=0$) or
\begin{equation}
\frac{\Delta r}{r}=\pm \frac{ \xi_o}{2a}-s_p \!\  \frac{\xi}{2a} \cos(3\theta_c),
\end{equation}
positive sign for electrons, negative for holes. There is always a radial shrinking in correspondence of self-trapped holes  in nanotubes with $p=2$, and radial expansion for electrons in nanotubes with $p=1$, as partially confirmed by density functional findings of Ref.~\cite{Verissimo0}.

While the casuistic of excitons is too rich to be dealt with here~\cite{Dresselhaus2}, we can still offer a few considerations for the $E_{11}$ case. As couplings of electron and holes with anisotropic deformations $\gamma_o$ cancel each other, we must take $\nu'=0$ and thus $\xi_o=0$ in our equations. Hence, we can have optomechanical effects of both elongation or contraction, depending on the chiral indices, as predicted by~(\ref{roxana}) with $\xi_o=0$. 
%Of course a full study of the entire excitonic spectra is needed to make sure that effects of different bands do not cancel. 

The treatment above applies to static polarons in not too small nanotubes, as we neglect the breaking of honeycomb symmetry brought up by the chiral vector.  That would introduce new terms in the elastic energy and result in curvature corrections  for very small radii~\cite{Nisoli}. Also, for small radii, orbital hybridization imposes a more sophisticate tight binding treatment. For the propagating polaron, another symmetry breaking comes from the velocity  vector, directed along the nanotube axis. Even for large nanotubes, we expects corrections (at lowest order quadratic) in the speed of the traveling polaron.~\cite{NisoliPRB}.

In conclusion, we have calculated the elastic deformations and electronic structure  of self-trapped electronic states in single wall carbon nanotubes. We found that elongation and torsion depends highly non monotonically on the wrapping indices,  whereas the polaron length does not.  The inner displacement is always axial and changes orientation depending on the chiral indices: this suggests a chirality dependent bond lengthening/contraction and optical mode frequency shift.  Extension of this model to excitons is particularly interesting.

We would like to thank Ryan Kalas (Los Alamos National Laboratory) for reading the manuscript. 
This work was carried out under the auspices of the National Nuclear Security Administration of the U.S. Department
of Energy at Los Alamos National Laboratory
under Contract No. DE-AC52-06NA25396.

\end{document}